\documentstyle[12pt]{article}
\newcommand{\ds}{\displaystyle}
\newcommand{\la}[1]{\label{#1}}
\newcommand{\re}[1]{\ (\ref{#1})}
\newcommand{\nn}{\nonumber}
\newcommand{\ed}{\end{document}}
\newcommand{\be}{\begin{equation}}
\newcommand{\ee}{\end{equation}}
\newcommand{\ba}{\begin{eqnarray}}
\newcommand{\ea}{\end{eqnarray}}
\newcommand{\bb}{\begin{thebibliography}}
\newcommand{\eb}{\end{thebibliography}}
\newcommand{\bi}[1]{\bibitem{#1}}
\newcommand{\ct}[1]{\cite{#1}}

\newcommand{\bc}{boundary condition}
\newcommand{\bm}{bag model}
\newcommand{\cf}{confinement}
\newcommand{\cst}{constituent}
\newcommand{\con}{contribution}
\newcommand{\cu}{current}

\newcommand{\dc}{distance}
\newcommand{\dt}{distribution}
\newcommand{\ef}{effect}
\newcommand{\ex}{experiment}
\newcommand{\fu}{function}
\newcommand{\hn}{hadron}

\newcommand{\Hw}{However}
\newcommand{\s}{instanton}
\newcommand{\ia}{interaction}
\newcommand{\me}{matrix element}
\newcommand{\n}{nucleon}
\newcommand{\np}{nonperturbative}
\newcommand{\ptb}{perturbative}
\newcommand{\p}{proton}
\newcommand{\q}{quark}
\newcommand{\ssb}{spontaneous chiral symmetry breaking}
\newcommand{\st}{structure function}
\newcommand{\tp}{topological}
\newcommand{\vm}{vacuum}
\newcommand{\wf}{wave function}
\newcommand{\w}{where}

\newcommand{\cd}{condensate}
\newcommand{\qm}{quark model}
\newcommand{\pl}{particle}
\newcommand{\str}{structure}
\begin{document}
{\hfill     preprint JINR E2-94-103}

{\hfill     March, 1994}\\
\begin{center}
{\bf
DEFICIENCY OF THE GROSS - LLEWELLYN SMITH SUM RULE \\
AND \\
QCD VACUUM POLARIZATION EFFECT
}\\[5mm]
A. E. DOROKHOV
\footnote{
E-mail: \large { dorokhov@thsun1.jinr.dubna.su}}\\
{\small \it Joint Institute for Nuclear Research,} \\
{\small \it 141980, Dubna, Moscow region, Russia}\\[3mm]

{\it Submitted to Physics Letters B}\\[3mm]

{\bf Abstract} \\
\end{center}
 From the analysis of the recent CCFR Collaboration data for the \st\
$xF_3(x,Q^2)$
($0.015<x<0.65$ and $1.2\ GeV^2 < |Q^2| < 501\ GeV^2$) of the deep inelastic
neutrino - nucleon scattering we conclude that probably part of the nucleon
baryon number is due to the vacuum polarization effects.
\vspace{1.0cm}

	The deep inelastic lepton - nucleon scattering processes (DIS)
occurring at small distances characterize the internal \str\  of the elementary
\pl s. In the past few years new \ex\ data with high precision and in large
kinematic region has became available.

Recently the QCD analysis of the most precise data for the neutrino - \n\ DIS
\st\  $xF_3(x,Q^2)$ measured by the CCFR Collaboration at the FERMILAB collider
\ct{CCFR} has been performed \ct{KatSid}. This analysis results in the
estimation
of the Gross - Llewellyn Smith sum rule (GLSsr) \ct{GLSsr} in wide region of
$Q^2$, $2\ GeV^2<Q^2<500\ GeV^2$,
\be
GLS(Q^2)=\frac{1}{2}\int\limits_{0}^{1}
{\ds xF_3^{\bar\nu p+\nu p}(x,Q^2)\over\ds x} dx
\la{GLS}\ee
and reveals at the level of the statistical \ex al errors the \ef\ of the
discrepancy between the measurements of the GLSsr and the \ptb\ QCD
prediction:
\be
GLS_{QCD}(Q^2)=3[1-{\ds \alpha_s(Q^2)\over\ds
\pi}+O(\alpha_s^2(Q^2))+O(1/Q^2)].
\la{ptGLS}\ee

The deficiency, $\Delta GLS\equiv GLS_{QCD}-GLS_{exp}$, at the squared
momentum  transfer $Q^2=10\ GeV^2$ with four active flavors and
$\Lambda^{(4)}_{\bar{MS}}=213\ MeV$ is equal to:
\be
\Delta GLS(Q^2=10\ GeV^2) = 0.18\pm0.12(stat)
\la{DGLS}\ee
and decreases only logariphmically with the squared momentum transfer over all
\ex al accessible region up to $500\ GeV^2$.
We choose the reference scale at $Q^2=10\ GeV^2$ \w\ the data are most
statistical valuable \ct{CCFR} and \w\ the high twist \ef s and the target
mass corrections are negligible \ct{KatSid}. Moreover, it is this scale \w\
large helicity and flavor asymmetry of the \p\ sea is observed in the EMC
\ct{EMC} and NMC \ct{NMC} \ex s.

In the present letter we suggest the mechanism explaining the possible
violation of the GLSsr based on the \np\ QCD dynamics. Keeping in mind
the different \ex al and theoretical uncertainties in extracting the
value \re{DGLS} we will consider this number as an upper bound of the
\ef . The mechanism suggested is highly related to the one
violating the Ellis - Jaffe and Gottfried sum rules \ct{DKSpin}.

In the framework of the parton model the GLSsr for the \p\ \st\ $F_3(x,Q^2)$
corresponds to the conservation of the baryon number, $B$,
\be
{\ds 1\over\ds 3}
\int\limits_{0}^{1} [(u(x,Q^2)-\bar u(x,Q^2))+(d(x,Q^2)-\bar d(x,Q^2))]\ dx=
B(1- {\ds \alpha_s(Q^2)\over\ds \pi}).
\la{parGLS}\ee
The baryon charge operator in the \qm\ is defined by
\be
\hat B={\ds 1\over\ds 6}\int\limits_{0}^{1}
(\{u^+(\vec x),u(\vec x)\}_+ +
\{d^+(\vec x),d(\vec x)\}_+) d\vec x
\la{BCh}\ee
and the baryon number is related to the low - energy spin - averaged \me\ of
the isoscalar vector \cu\
$J_\mu(\vec x)=\bar u\gamma_\mu u +\bar d\gamma_\mu d$ over the \p\ state:
\be
<p|J_\mu(0)|p>=12p_\mu B.
\la{BN}\ee
If the \p\ state $|p_0>$ contained only free \q s, then the baryon number
would be equal one exactly, $B=1$. The index $0$ of $|p_0>$ means that
a \p\ (and \q s) is considered over perturbative QCD \vm\ with zero \con\
of Dirac sea \q s to the baryon number: $<p_0|\hat B^{sea}|p_0>=0$.

\Hw , the physical \p\ is immersed in the strong interacting medium and the
phenomena
of the confinement and of the \ssb\ occur. As it has been shown by Skyrme and
Witten \ct{Skyrme,Witten} this highly nonlinear QCD vacuum can carry its own
baryon number:
\be
B^{Skyrme}={\ds 1\over\ds 24\pi^2}\epsilon_{0\mu\lambda\rho}
\int Tr \{R_\mu R_\lambda R_\rho\} d\vec x,
\la{BSky}\ee
\w\ $R_\mu = (\partial_\mu U)U^+$ with $U^+U=1$ is constructed from bosonic
fields.  It is the \ef\ of the fermion - boson transmutation. In the Skyrme
model the chiral soliton baryon charge \re{BSky} is fully compensated for by
the negative baryon charge induced by sea \q s.

Later Rho, Goldhaber and Brown \ct{Rho} and Goldstone and Jaffe
\ct{GoldJaffe} have suggested that the baryon number \re{BN} of the \p\
surrounded by the nontrivial (Skyrme) \vm\ could be distributed
between the normal (canonical) \q\ contribution, $B^{valence}$, and the part
anomalously
induced by the \vm\ polarization,  $B^{sea}$:
\be
B=B^{valence}+B^{sea}.
\la{BNChBM}\ee
The latter is related to the influence of  the regularization procedure on
the symmetry properties of the theory and is of pure quantum origin. Within
the chiral bag model for the physical \p\ state $|p>$ the valence and
sea polarization parts of the baryon number are equal to:
\ba
&&<p|\hat B^{valence}|p>=1, \nn\\
&&<p|\hat B^{sea}|p>=-B^{Skyrme},
\la{BNvs}\ea
correspondingly.
We can write the following sum rule:
\be
B^{valence} + B^{sea} + B^{Skyrme} = 1,
\la{Bsr}\ee
with $B^{sea} = -B^{Skyrme}$ (by definition) and $B^{Skyrme}$ is invisible
in DIS due to its
bosonic origin. This interpretation of the anomalous sea \q\ \con\ to the
baryon charge is in complete analogy with the interpretation of the total
angular momentum sum rule for the proton \ct{BrElKar}. There, the relative
angular momentum inactive in DIS is produced to compensate for the negative
helicity of sea \q s created in the field of strong \vm\ fluctuation, \s\
\ct{DKSpin}.

In the framework of the chiral \bm\ \ct{ChBM} when a massless
Dirac \q\ field is confined to a finite region of space by means of a chiral
\bc\ parametrized by a chiral angle $\Theta$
characterizing a leakage of the baryon charge,
the anomalous baryon number of
the vacuum is equal to \ct{GoldJaffe}
\ba
&&B^{Skyrme}(\Theta)=
-{\ds 1\over\ds \pi}
(\Theta- {\ds 1\over\ds 2} \sin{ 2\Theta}),\ \ \ \ \ \ \ \ \
-{\ds \pi\over\ds 2} < \Theta < {\ds \pi\over\ds 2},  \la{Ban}\\
&&B^{Skyrme}(\Theta+\pi)=B^{Skyrme}(\Theta),\ \ \ \ \ \ \
\mbox{outside the interval}\ \ \ \
 [-{\ds \pi\over\ds 2},{\ds \pi\over\ds 2}].\nn
\ea
This expression is given for the boundary separating the region of
intermediate and large \dc s where soft \vm\ \ef s occur \tp ly
equivalent to a sphere. The chiral \bc\ of general form
\be
-i(\hat n\vec\gamma)\ q_L|_S=M(\Theta)\ q_R|_S,
\la{ChBC}\ee
\w\ $\hat n$ is the outward normal to the surface, is due to specific condition
of the \cf\ of \q s in the closed region. Matrix $M$ is such that the axial
vector isotriplet \cu\ conservation should be satisfied and simultaneously
the flavor singlet axial \cu\ should have the anomaly. These requirements
fix the form of the chiral \bc\ as an \ef ive surface \ia\ of
the \q\ fields confined to the \hn\ with external fields from the \vm\ \cd\
due to \s\ exchange \ct{DKZbm}:
\be
-i\vec\gamma\cdot \hat n\ q|_s=
\exp[i\gamma_5 \Theta(\vec\tau\cdot \hat n + 1)]\ q|_s.
\la{ChBC1}\ee

As it has been shown in \ct{GoldJaffe} we have the following picture of the
baryon charge leakage. The chiral angle $\Theta$ varies from zero at very
large value of bag radius to $-\pi$ as bag radius goes to zero. It corresponds
to change of the baryon charge carried by Dirac sea \q s from zero at chiral
angle $\Theta=0$ (large $R$) to -1 at $\Theta=-\pi$  $(R=0)$. When $\Theta$
pass $-\pi/2$ the occupied positive \q\ mode transit sharply into a negative
- charge level and the baryon charge of the Dirac sea changes by $-1$
\re{Ban}.

Now we can relate the deficiency of the GLSsr \re{DGLS} with the anomalous
\vm\ baryon number
\be
-{\ds 1\over\ds \pi}(\Theta - {\ds 1\over\ds 2}\sin 2\Theta) =
0.06\pm0.04
\la{ExDGLS}\ee
and then estimate the value of the chiral angle:
\be
\Theta = -{\ds \pi\over\ds 4}\ \left(0.86
\begin{array}{l}
+0.18 \\
-0.27
\end{array}
\right).
\la{ExTh}\ee
The numbers \re{ExDGLS} and \re{ExTh} correspond to an upper bound of
the \ef .

At the same time the isovector chiral flow through the surface controlled
by the \bc\ \re{ChBC1} is zero due to the equal number of left - and right -
handed chiral \q s. The pseudoscalar isosinglet coupling \re{ChBC1} at the
surface has a consequence on the description of the flavor singlet \cu\
of the \p\  (\p\ spin) \ct{DKZbm} and leads to the color anomaly \ct{ColAn}.

Thus we can interpret the possible violation of the Gross - Llewellyn Smith sum
rule observed by CCFR Collaboration in neutrino - nucleon DIS in wide $Q^2$
interval as a hint at a
large polarization \ef\ in the \np\ QCD \vm\ surrounding the \hn .
We also stress that the peculiar \ia\ of the \cst s induced by \s s is also
responsible for large helicity and flavor asymmetry of the sea \q s in the \p\
\wf\ and the sea \q\ \dt\ \fu s.
These and related questions are currently under investigation.

In addition, the \ex al investigation
(and theoretical understanding) of the behavior of the structure function
$F_3(x,Q^2)$ in the region of small $x$ and the calculation of the different
QCD corrections at large $Q^2$ is necessary. The consideration of the
nuclear \ef s is also important to have an unambiguous conclusion about the
value of the GLSsr breaking.

I am thankful to A.V. Sidorov for the stimulating discussions and informing me
about the CCFR Collaboration results
and P.N. Bogolubov, S.B. Gerasimov, A.L. Kataev, N.I. Kochelev and A.W. Thomas
for discussions.

\bb{99}
\bi{CCFR} CCFR Collab. {\it Phys. Lett.} {\bf B317} (1993) 665;
{\it Phys. Rev. Lett.} {\bf 71} (1993) 1307.
\bi{KatSid} A.L. Kataev, A.V. Sidorov {\it CERN preprint}
CERN-TH.7160/94 / JINR E2-94-45, February, (1994),
{\it Phys. Lett. B submitted}.
\bi{GLSsr} D.J. Gross, C.H. Llewellyn Smith
{\it Nucl.\ Phys.\ }{\bf B14} (1969) 337.
\bi{EMC} EMC, J. Ashman et al. {\it Phys.\ Lett.\ }{\bf B206} (1988)
364; {\it Nucl.\ Phys.\ }{\bf B328} (1989) 1.
\bi{NMC} NMC, D. Allasia et al, {\it Phys.\ Lett.\ }{\bf B249} (1990) 366;
P. Amaurdruz et al, {\it Phys. Rev. Lett.\ }{\bf 66} (1991) 2712.
\bi{DKSpin} A. E. Dorokhov, N. I. Kochelev {\it Mod.\ Phys.\ Lett.\ }{\bf A5}
 (1990) 55; {\it Phys.\ Lett.\ } {\bf B259} (1991) 335.
\bi{Skyrme} T.H.R. Skyrme {\it Proc. Roy. Soc.} London, {\bf A260} (1961) 127;
{\it Nucl. Phys.} {\bf 31} (1962) 556.
\bi{Witten} E. Witten {\it Nucl. Phys.} {\bf B223} (1983) 422, 433.
\bi{Rho} M. Rho, A.S. Goldhaber, G.E. Brown {\it Phys. Rev. Lett.}
{\bf 51} (1983) 747.
\bi{GoldJaffe} J. Goldstone, R.L. Jaffe  {\it Phys. Rev. Lett.} {\bf 51}
(1983) 1518.
\bi{ChBM} C.G. Callen, R.F. Dashen, D.J. Gross
{\it Phys. Rev.} {\bf D19} (1979) 1826;
G.E. Brown, M. Rho, V. Vento {\it Phys. Lett.} {\bf 84B} (1979) 383;
R.L. Jaffe, in {\it Pointlike Structure Inside and Outside the Nucleon},
Proceedings of the 1979 Erice Summer School "Ettore Majorana" edited by
A. Zichichi (Plenum, New York, 1981);
A.W. Thomas, S. Theberge, G.A. Miller {\it Phys. Rev.} {\bf D24} (1981) 216.
\bi{BrElKar} S.J. Brodsky, J. Ellis, M. Karliner
{\it Phys.\ Lett.\ } {\bf B206} (1988) 309;
S. Forte, {\it Phys.\ Lett.\ }{\bf B224} (1989) 189.
\bi{DKZbm} A. E. Dorokhov, N. I. Kochelev, Yu. A. Zubov {\it Sov. J. Part.
Nucl. Phys.}
{\bf 23} (1992) 522 (1192).
\bi{ColAn} H.B. Nielsen, M. Rho, A. Wirzba, I. Zahed {\it Phys. Lett.}
{\bf B269} (1991) 389.
\eb
\ed